\documentclass[epjCONF]{svjour}
\usepackage{graphicx}
\usepackage[varg]{txfonts} 
\usepackage[latin1]{inputenc}
\session-title{%
19$^{\textnormal{\footnotesize th}}$ International %
IUPAP Conference on Few-Body Problems in Physics%
}
\begin{document}
\title{%
Electromagnetic interactions in Halo Effective Field Theory
}%
\author{%
R. Higa\inst{1,2}\fnmsep\thanks{\email{R.Higa@rug.nl}} 
}
\institute{%
Helmholtz-Institut f\"ur Strahlen- und Kernphysik, Universit\"at Bonn, 
Nu\ss allee 14-16, 53115 Bonn, Germany
\and %
Kernfysisch Versneller Instituut, Rijksuniversiteit Groningen,
Zernikelaan 25, 9747 AA Groningen, The Netherlands
}
\abstract{
After a brief discussion of effective field theory applied to nuclear 
clusters, I concentrate on the inclusion of two particular aspects, namely, 
narrow resonances and electromagnetic interactions. 
As examples of applications, I present the details of our studies on 
alpha-alpha and proton-alpha scattering. 
} 
\maketitle
%
%
%
\section{Introduction}
\label{SchmidtPL_intro}


Two-particle scattering at low enough energies are insensitive to 
the details of the interparticle interaction, provided that the latter 
has a limited range of action. 
The intuitive argument behind this general property is that momenta whose 
wavelengths are much larger than the interaction range cannot resolve its 
finer details. 
The elastic amplitude incorporates unitarity constraints and an universal 
low-energy expansion, {\em a.k.a.} the effective range expansion (ERE), 
whose coefficients encode the low-energy behavior of the forces for a 
particular system. 
This simple observation, made by Bethe and others~\cite{bethe49} 
more than fifty years ago, became a benchmark for theorists trying to 
model the dynamics of the interaction. 

The same line of reasoning is shared by effective field theories (EFT), 
where only the relevant low-energy degrees of freedom are explicitly 
taken into account. Modes that are active only at high energies are 
``frozen'' into the low-energy constants. 
In the particular case of EFT with short-range interactions, 
the two-body scattering amplitude is equivalent to a low-energy 
expansion of the ERE amplitude~\cite{biraere}. 
However, the EFT approach has extra 
features over the effective range theory. First, the systematic 
expansion of the amplitude allows for a reliable and unbiased estimate 
of the theoretical uncertainties. Second, symmetries of the problem or the 
way they are broken can be implemented in a straightforward way (for 
instance, electromagnetic and weak currents, isospin, parity). 
And third, the ability to extend the formalism to systems with more 
than two particles. 

One example of the second feature illustrates a difference between the 
EFT and ERE approaches. In the process $np\to d\gamma$, the ${\rm M1_v}$ 
transition contains at NLO a four-nucleon-photon operator that 
cannot be generated by the ERE approach \cite{CRS}. The associated 
coupling can be fixed from cold neutron capture rate at incident 
neutron speed of 2200 m/s. 
Rupak improved the calculation of ${\rm M1_v}$ and ${\rm E1_v}$ 
transitions to ${\rm N^2LO}$ and ${\rm N^4LO}$, respectively, 
lowering the theoretical uncertainty in the $np\to d\gamma$ reaction 
to about 1\% for center-of-mass energies up to 1 MeV \cite{gautam}. 

When extended to three or more particles with large scattering length(s), 
EFT provides new insights 
and explains certain universal properties that are independent of 
the interaction details. For instance, the related Thomas and Efimov 
effects~\cite{tho-efi} are intimately connected to the behavior 
of the leading three-body counterterm under variations of the 
momentum cutoff in the dynamical equations. Such counterterm 
is necessary to properly renormalize the theory, as well as to guarantee 
a unique solution in the limit when the cutoff is taken to infinity 
\cite{eftrev1,eftrev2}. It exhibits a limit cycle behavior, that is, 
a log-periodic dependence with the cutoff, that explains the 
geometrically-separated bound states in the Efimov spectrum. 
For nuclear systems with three and four nucleons, the theory provides 
a convincing explanation to certain correlations, like the almost 
linear dependence of the spin-doublet neutron-deuteron scattering 
length with the triton binding energy~\cite{BHvK00,platter06} or 
similar dependence of the latter with the alpha particle binding 
energy~\cite{PHM05}, known respectively as the Phillips and Tjon lines. 

Due to the universal character, effective field theories with short-range 
interactions have been successfully applied to distinct areas of physics 
---atomic, particle, and few hadron systems. For a more complete overview, 
see Refs.~\cite{eftrev1,platterrev}. 
Here I will concentrate on applications to nuclei that behave as 
systems of loosely bound clusters, which also include the nowadays 
popular halo nuclei. In Section~\ref{sec:haloEFT} the general features 
of the so-called halo/cluster EFT are presented in certain detail, 
with emphasis on aspects quite often faced when dealing with nuclear 
clusters. Sections~\ref{sec:aa} and \ref{sec:pa} present applications 
of halo/cluster EFT to alpha-alpha ($\alpha\alpha$) and proton-alpha 
($p\alpha$) scattering, 
and in Section~\ref{sec:end} are the concluding remarks. 

\section{EFT for nuclear clusters}\label{sec:haloEFT}

The relevant degrees of freedom in halo/cluster EFT are the 
structureless, weakly bound objects represented by respective 
field operators. In the case of halo nuclei, these are the core 
nucleus and the valence nucleons. Effects like nucleon excitations 
inside the core, pion or nucleon exchanges, take place at energies 
well above the binding energy that holds the clusters together. 
The former has an associated momentum scale $M_{hi}$ of the order of 
the inverse of the interaction range, while in the latter case 
the momentum scale $M_{lo}\ll M_{hi}$ is inversely proportional to 
the size of the halo system. 
As explained in the following, this separation of scales is, from the 
EFT point of view, the outcome of a fine-tune in the coupling constants, 
similar to the one responsible for generating a shallow bound state in 
the spin-triplet nucleon-nucleon channel. 
In fact, due to its large extension compared to the typical range of the 
proton-neutron interaction, the deuteron can be seen as the simplest halo 
nucleus, with a loosely bound neutron surrounding a proton core. 

In order to understand the formalism, let us consider a system of 
identical bosons, with mass $m_{\alpha}$, represented by a field $\phi$ 
and interacting via a pairwise $S$-wave short-range force. The most 
general Lagrangian respecting the relevant symmetries of the system 
(parity, total angular momentum, approximate isospin and non-relativistic 
Galilean invariance) is given by 
\begin{eqnarray}
{\cal L}&=&
\phi^{\dagger}\Bigg[i\partial_0+\frac{\vec\nabla^2}{2m_{\alpha}}\Bigg]\phi
-\,d^{\dagger}\Bigg[i\partial_0+\frac{\vec\nabla^2}{4m_{\alpha}}-\Delta\Bigg]d
\nonumber\\&&
+g\,\Big[d^{\dagger}\phi\phi+(\phi\phi)^{\dagger}d\Big]
+\cdots\,,
\label{eq:LOLag}
\end{eqnarray}
where we introduce an auxiliary (dimeron) field $d$, 
with ``residual mass'' $\Delta$, 
carrying the quantum numbers of two bosons in $S$-wave 
and coupling with their fields through the coupling constant $g$. 
The dots stand for higher order terms in a derivative expansion. 
The building blocks are the boson and dimeron propagators, the LO 
coupling $g$ between the dimeron and two boson fields, and higher-order 
couplings that can be rewritten, via field redefinitions, as powers of 
the dimeron kinetic energy. The boson and dimeron propagators read 
\begin{eqnarray}
iS_{\alpha}(q_0;\mathbf{q})&=&\frac{i}{q_0-\mathbf{q}^2/2m_{\alpha}
+i\epsilon}\,,
\\
iD(q_0;\mathbf{q})&=&-\frac{i}{q_0-\mathbf{q}^2/4m_{\alpha}
-\Delta+i\epsilon}\,,
\label{eq:dimprop}
\end{eqnarray}
where the precise form of the latter depends on the magnitude of each term 
in the denominator. 
The power counting specifies the amount of fine-tuning in the low-energy 
couplings, responsible for assigning a well-defined order to 
every possible Feynman diagram that contributes to the scattering 
amplitude. 
In a natural scenario, where fine-tuning is absent, the effective 
couplings in Eq.(\ref{eq:LOLag}) depend only on the high-energy scale 
$M_{hi}$ of the theory. That corresponds to 
$\Delta\sim M_{hi}^2/m_{\alpha}$ and $g^2\sim M_{hi}/m_{\alpha}^2$.
The bare dimeron propagator (\ref{eq:dimprop}) is static at leading order 
(LO), 
\begin{displaymath}
iD(p_0;\mathbf{p})\simeq
\frac{-i}{-\Delta+i\epsilon}\sim O(m_{\alpha}/M_{hi}^2)\,.
\end{displaymath}
The LO tree diagram is proportional to 
$a_0=m_{\alpha}g^2/4\pi\Delta\sim O(1/M_{hi})$, where $a_0$ is the 
boson-boson scattering length. The one-loop graph receives two additional 
factors from the coupling constant $g$, one dimeron propagator, one-loop 
integration measure $dq_0\,d^3q\sim k^5/m_{\alpha}$, and two intermediate 
boson propagators $S_{\alpha}^2\sim (m_{\alpha}/k^2)^2$. The net result is 
a suppression of $k/M_{hi}$ compared to the tree graph, with $k$ the 
typical low-energy momentum under consideration. The next order, 
$k^2/M_{hi}^2$ smaller than the tree level, comprises two-loop 
contributions proportional to $(a_0k)^2$ 
and one kinetic term insertion to the dimeron propagator proportional to 
$a_0r_0k^2$, where $r_0=4\pi/m_{\alpha}^2g^2\sim 1/M_{hi}$ is the 
boson-boson effective range. 
The amplitude then amounts to a simple Taylor expansion in powers of 
the expansion parameter $k/M_{hi}$.

However, a natural perturbative treatment is unable to account for 
a plethora of nuclear phenomena, where bound states and resonances 
are the rule rather than exceptions. That demands a certain amount 
of fine-tuning in the parameters of the effective Lagrangian. The physics 
of shallow $S$-wave bound states can be described by fine-tuning the 
parameter $\Delta$ in Eq.(\ref{eq:LOLag}), responsible for driving the size 
of the scattering length $a_0$. Setting $\Delta\sim M_{hi}M_{lo}/m_{\alpha}$ 
leads to a large $a_0\sim 1/M_{lo}$, with $M_{lo}$ the new momentum scale 
associated to the fine-tuning. This situation is analogous to the 
nucleon-nucleon ($NN$) case \cite{eftrev1,eftrev2}. 
A consequence of this new scale is that higher-loop graphs built up 
from the LO boson-boson-dimeron coupling $g$ are no longer suppressed 
when $k\sim M_{lo}$, 
but contribute at the same order as the lower-loop and tree graphs. 
The resummation of this class of diagrams leads to a LO amplitude in 
the form of a truncated 
effective range expansion up to its first coefficient. The scattering 
length then becomes the only relevant scale and (for positive values) 
predicts the existence of a shallow bound state 
$E_B\approx 1/(m_{\alpha}a_0^2)$. 
The renormalization group (RG) analysis of the Schr\"odinger equation 
demonstrates that two-body systems with large scattering length consist of 
strongly interacting systems with an RG-flow towards a non-trivial fixed 
point \cite{birse}, in contrast to the natural case of weakly interacting 
particles, where the RG flows towards a stable, trivial fixed point. 

When extended to three-body systems, this EFT provides new insights on 
some universal features, as mentioned in the introduction. 
An investigation searching for these universal effects in two-neutron 
halos was performed in Ref.~\cite{CH1}, 
where it was found the ${}^{20}$C nuclei as possible candidate to 
have an Efimov-like spectrum (see also~\cite{amorimetal}).
Non-universal (effective range) corrections have recently being incorporated 
to this study~\cite{CH2}. 

There is an additional phenomenum quite often present in nuclear 
clusters, namely, the existence of low-energy resonances in the 
continuum. One could then expect the need for extra amount of 
fine-tuning. This is the case for neutron-alpha ($n\alpha$) scattering 
\cite{BHvK1,BHvK2}, where the presence of a narrow resonance in the $P_{3/2}$ 
channel around $E_n\approx 1$ MeV requires adjustment of 
the coupling constants of the theory to enhance the $P_{3/2}$ amplitude 
a few orders of magnitude away from the natural assumption. 
Another example is the $\alpha\alpha$ system \cite{HHvK}, which contains 
an $S$-wave resonance about 0.1 MeV above the elastic threshold. 
The $\alpha\alpha$ scattering amplitude can be described by the 
Lagrangian (\ref{eq:LOLag}), but assuming $\Delta\sim M_{lo}^2/m_{\alpha}$. 
This is necessary to generate an even larger $\alpha\alpha$ scattering 
length, $a_0\sim M_{hi}/M_{lo}^2$. These scalings will be discussed in 
detail in the next section. 
With this power-counting the dimeron propagator is no longer static, as the 
kinetic and residual mass terms are now of comparable order. In momentum 
space it has the form of Eq.(\ref{eq:dimprop}).
The presence of the kinetic term in the dimeron propagator in practice 
resums effective range $r_0$ corrections to all orders. 
This resummation is necessary to reproduce a narrow resonance at 
low energy\cite{HHvK,BHvK1,BHvK2}. 

\begin{figure}[htb]
\centerline{
\includegraphics[width=2.5in]{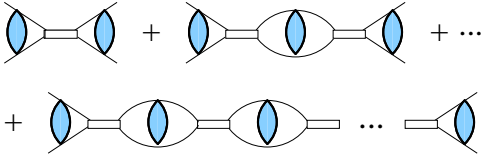}
}
\vspace*{8pt}
\caption{Graphic representation of $T_{CS}$, as sum of graphs with multiple 
insertions of the bare dimeron propagator (double line) and the ``Coulomb 
bubble loop". The latter contains intermediate Coulomb photons resummed to 
all orders (shaded ellipse).
\protect\label{fig1}}
\end{figure}
For the case of charged bosons, electromagnetic interactions can be 
incorporated into Eq.(\ref{eq:LOLag}) via the usual minimal substitution. 
Among them, Coulomb photons are the dominant ones at low energies \cite{HHvK}. 
Coulomb interactions were formulated in the EFT framework by Kong and 
Ravndal~\cite{clbeft} for the two-proton system, and can be extended in a 
straightforward way to include resonances \cite{HHvK}. The idea 
relies on the two-potential formalism, where the amplitude is separated 
in a term that contains only the two-particle pure Coulomb scattering $T_C$, 
plus the Coulomb-modified strong amplitude $T_{CS}$. The latter is 
diagramatically illustrated by Fig.~\ref{fig1} (see caption) and has the 
form of a geometric series. Like in the proton-proton case~\cite{clbeft}, 
the power counting for narrow resonances demands its resummation. 
Up to next-to-leading order (NLO) one gets for our two-boson example, 
with reduced mass $\mu$ and charge $Z_{\alpha}$, 
\begin{eqnarray}
T_{CS}&=&-\frac{2\pi}{\mu}\,C_{\eta}^2\,e^{2i\sigma_0}\,\Bigg[
\frac{1}{-\frac{1}{a_0}\!+\!k^2\,\frac{r_0}{2}\!-\!2k_C H} 
\nonumber\\&&
+
\frac{{\cal P}_0}{4}
\frac{k^4}{(-\frac{1}{a_0}\!+\!k^2\,\frac{r_0}{2}\!-\!2k_C H)^2} 
\Bigg]\,,
\label{eq:ampcs2}
\end{eqnarray}
where 
\begin{equation}
k_C=\frac{\mu Z_{\alpha}^2e^2}{4\pi}=\mu Z_{\alpha}^2\alpha_{em}
\label{eq:kcdef}
\end{equation}
is the inverse of the Bohr radius, and ${\cal P}_0$, the shape parameter. 
The strength of Coulomb photons is driven by the Sommerfeld parameter 
\begin{equation}
\eta(k)=\frac{k_C}{k}\,,
\end{equation}
and the Coulomb-modified strong amplitude is modulated by the Sommerfeld 
factor, 
\begin{equation}
C_{\eta}^2=\frac{2\pi\eta}{e^{2\pi\eta}-1}\,,
\end{equation}
which quantifies the probability of finding the two bosons at the same 
point in space. 
The phase of pure Coulomb relative to free scattering, for a given partial 
wave $L$, reads 
\begin{equation}
\sigma_L=\arg\Gamma(L+1+i\eta)
=\frac{1}{2i}\ln\left[\frac{\Gamma(L+1+i\eta)}{\Gamma(L+1-i\eta)}\right]\,.
\end{equation}
The last Coulomb ingredient in Eq.(\ref{eq:ampcs2}) is the function 
\begin{equation}
H(\eta)=\psi(i\eta)+(2i\eta)^{-1}-\ln(i\eta)\,,
\label{eq:Hdef}
\end{equation}
and explicitly shows the complicated analytic structure of Coulomb 
scattering at low energies. 
Nevertheless, apart from the Sommerfeld factor and the $S$-wave 
Coulomb phase, one notices that the effect of dressing the strong 
amplitude with non-perturbative Coulomb photons 
amounts to, apart from a multiplicative factor in the numerator, 
replacing the unitarity term $-ik$ by $-2k_C\,H(\eta)$ 
in the denominator of the amplitude. Similar expression can be derived 
for $P$-wave interactions and will be presented in Section~\ref{sec:pa}.

\section{$\alpha\alpha$ scattering}\label{sec:aa}

\subsection{power counting: pinning the scales down}

In order to better understand the scalings of the EFT parameters in 
the presence of a resonance, it is helpful to provide some numbers for 
the relevant physical quantities. It is well-known that the $\alpha\alpha$ 
system is dominated by $S$-wave at low energies \cite{aaa69}, 
having a very narrow resonance at $E_{R}\simeq 92$~keV 
and width $\Gamma_R\approx6$~MeV. To this resonance is commonly assigned 
the ${}^{8}$Be ground state. 
The $\alpha\alpha$ scattering length $a_0\sim 2000$~fm is about 
three orders of magnitude larger than the $\alpha$ matter radius, 
suggesting the large amount of fine-tuning discussed in the last section. 
In contrast, the effective range $r_0\approx 1$ fm and shape parameter 
${\cal P}_0\approx -1.7 {\rm fm}^3$ obey the expected natural scalings 
of $1/M_{hi}$ and $1/M_{hi}^3$, respectively. 
It seems therefore reasonable to start with the scalings 
$\Delta\sim M_{lo}^2/\mu$ and $g^2/2\pi\sim M_{hi}/\mu^2$. 
With the evaluation of Feynman diagrams and after the required 
resummation described in the previous section one finds 
\begin{equation}
T_{CS}=-\frac{2\pi}{\mu}C_{\eta}^2e^{2i\sigma_0}
\left[-\frac{2\pi\Delta^{(R)}}{\mu g^2}+\frac{\pi}{\mu g^2}k^2
-2k_CH(\eta)\right]^{-1}+\cdots
\end{equation}
where $\Delta^{(R)}$ is the only parameter in the Lagrangian, up to NLO, 
that is renormalized by the Coulomb loops, 
\begin{eqnarray}
\Delta^{(R)}&=&
\Delta (\kappa)
+\frac{\mu g^2}{2\pi}
\Bigg\{
\frac{\kappa}{D-3}
\nonumber\\[1mm]&&
+2k_C\left[\frac{1}{D-4}
-\ln\left(\frac{\sqrt{\pi}\kappa}{2k_C}\right)-1+\frac{3}{2}\,C_E\right]
\Bigg\}\,,
\label{eq:aarenorm}
\end{eqnarray}
with $D$ the number of space-time dimensions, $\kappa$ the renormalization 
scale parameter, and $C_E=0.577\ldots$ the Euler-Mascheroni constant. 
The amplitude matches with the expansion of the ERE amplitude 
(\ref{eq:ampcs2}), from where one identifies the ERE parameters in terms 
of the EFT couplings. One then finds $a_0\sim M_{hi}/M_{lo}^2$. 
The low-energy scale $M_{lo}\sim\sqrt{m_{\alpha}E_R}\approx 
20$~MeV is nearly seven times smaller than a high momentum scale 
associated to either the pion mass or the excitation energy of the 
alpha particle, $M_{hi}\sim m_{\pi}\sim\sqrt{m_{\alpha}E^{*}_{\alpha}} 
\approx 140$~MeV. 
Based on these numbers, one could expect convergent 
results for observables at laboratory energies up to 3~MeV. 

The Bohr momentum $k_C$ provides the scale of Coulomb interactions. 
Due to the large $\alpha\alpha$ reduced mass in its definition 
(\ref{eq:kcdef}), $k_C$ turns out to be numerically of the order of 
$M_{hi}$. 
One can therefore expects large and important electromagnetic 
contributions. 
Nevertheless, it is interesting to discuss the other limit $k_C\to 0$, 
when Coulomb is turned off. In this case one has 
$2k_CH(\eta)\to ik$ and the denominator of the LO amplitude reads 
\begin{equation}
T_{\rm LO}^{-1}\propto -1/a_0+r_0k^2/2-ik\,.
\end{equation}
For momenta $k\sim M_{lo}$ the first two terms are suppressed by 
$M_{lo}/M_{hi}$ compared to the last one. Therefore, all that is left at 
LO is the unitary term $1/(-ik)$. 
In this limit, the ${}^8$Be system exhibits non-relativistic conformal 
invariance \cite{msw} and the corresponding three-body system, 
${}^{12}$C, acquires an exact Efimov spectrum \cite{eftrev1,BHvK00}. 
If it was possible to somehow shield the charges of the $\alpha$ 
particles, the above limit strongly suggests that two and three chargless 
$\alpha$ particles would probably be the best nuclear systems to observe 
few-body universality. 
The above scenario certainly changes for a physical value of $k_C$, 
due to the breaking of conformal invariance by the $1/r$ Coulomb force. 
Nevertheless, the fact that the ground state of ${}^{8}$Be and the Hoyle 
state in ${}^{12}$C remain very close to the threshold into 
$\alpha$-particles suggests that this conformal picture is not far 
from the real case. 

The breaking of scale invariance by the Coulomb interaction 
introduces the scale $k_C$ in the propagation of two charged particles. 
As we have seen in Eq. (\ref{eq:ampcs2}), the unitarity term is modified. 
The balance between strong-interaction terms and Coulomb-modified 
propagation now depends on both the strong-interaction scale $M_{lo}$ 
and $k_C$. While the former is $k_R=\sqrt{m_\alpha E_R}\sim 20$~MeV, 
the latter is $k_C=\alpha_{em} Z_\alpha^2 \mu \sim 60$~MeV, and therefore 
a relative strength of $k_C/k_R\sim 3$. 
For momenta $k\sim k_R$, we are clearly in the deep non-perturbative 
Coulomb region. 
The Sommerfeld parameter $\eta$ reaches large values and the function 
$2k_C H(\eta)$ is very different from the usual unitarity term $ik$. 
Instead of hampering any simplification, it actually allows a low-energy 
expansion of the Coulomb function $H$. 
Using Stirling's series,
\begin{equation}
\ln\Gamma(1+z)= \frac{1}{2}\,\ln 2\pi+\left(z+\frac{1}{2}\right)\ln z
-z+\frac{1}{12z}-\frac{1}{360z^3}+\cdots\,,
\end{equation}
and $\psi(z)\equiv (d/dz)\ln\Gamma(z)$ in Eq. (\ref{eq:Hdef}) gives 
\begin{equation}
H(\eta)
= \frac{1}{12\eta^2}+\frac{1}{120\eta^4}+\cdots
+\frac{i\pi}{e^{2\pi\eta}-1}\,.
\label{eq:Happrox}
\end{equation}
The unitarity term is thus replaced by $2k_C H(\eta)\sim k^2/6k_C$ at LO. 
This is now a factor $k/6k_C$ smaller in magnitude 
than the unitarity term in the absence of Coulomb, 
and comparable to the effective-range term coming from 
the dimeron kinetic term. 
One can grasp it automatically if one considers $3k_C \sim M_{hi}$, 
as it appears to be the case numerically. 

\subsection{resonance pole expansion}
\label{sec:aa-pole}

Before incorporating the simplifications on the function $H(\eta)$, one 
should take a careful look at the EFT amplitude up to NLO. First, an 
isolated remark. Since 
we assume that ${\cal P}_0$ is a parametrically small correction, 
one is allowed to reshuffle the perturbative series and resum 
${\cal P}_0$ to all orders, thus recovering the ERE formula,
\begin{equation}
T_{CS}^{\rm (ERE)}=-\frac{2\pi}{\mu}\,\frac{C_{\eta}^2\,e^{2i\sigma_0}}
{-1/a_0\!+\!k^2\,r_0/2\!\!-k^4\,{\cal P}_0/4\!-\!2k_C H}\;.
\label{eq:ere}
\end{equation}
Second, one notices that equation (\ref{eq:ampcs2}) is valid for generic 
momenta $k\sim M_{lo}$. However, it fails in the immediate proximity of $k_R$. 
This situation is familiar from the neutron-alpha case \cite{BHvK2}. 
The power counting works for $k\sim M_{lo}$ 
except in the narrow region $|k-k_R|= {\cal O}(M_{lo}^2/M_{hi})$ 
where the LO denominator approaches zero and a resummation of the NLO 
term, here associated with the shape parameter, is required. 
As one gets closer to the resonance momentum $k_R$, higher-order terms 
in the ERE are kinematically fine-tuned as well. 
This happens because the imaginary part of the denominator 
is exponentially suppressed by a factor $\exp(-2\pi\eta_R)\sim 10^{-8}$ 
and the real part is allowed to be arbitrarily small. 
Nevertheless, this kinematical fine-tuning is not a conceptual problem. 
From the EFT point of view, each new fine-tuning can be accommodated by 
reshuffling the series and redefining the pole position. Such a procedure 
works fine with a small number of kinematical fine-tunings, but is not 
practical in the $\alpha\alpha$ system. A better alternative is to 
perform an expansion around the resonance pole position, starting from 
the resummed (ERE) amplitude. 
The situation here is nearly identical to the $NN$ system, where one can 
choose to expand the amplitude around the bound-state pole \cite{pole} 
rather than around zero energy. 

A great simplification is achieved from the fact that the resonance lies 
in the deep Coulomb regime, 
where Eq.~(\ref{eq:Happrox}) provides an accurate representation of 
$H$ up to the precision we are considering. 
The real terms shown in Eq.~(\ref{eq:Happrox}) 
then become an expansion in powers 
of $\sim (k/3k_C)^2={\cal O}(k^2/M_{hi}^2)$.
Given the asymptotic feature of the Stirling's series, 
at some point the remaining terms can no longer be expanded; at that point 
the remainder should be treated exactly. 
In lowest orders, however, we can use the successive terms shown in 
Eq. (\ref{eq:Happrox}): 
numerically, the terms up to $\eta^{-4}$ 
work to better than 3\% for $E_{LAB}=3$~MeV. 

The expansion (\ref{eq:Happrox}) not only simplifies a complicated 
function of $k_C/k$, but also makes the physics around the resonance 
more transparent. 
Since the ``size'' of the resonance, $1/k_R$, is much larger 
than the Bohr radius $1/k_C$, the Coulomb interaction is effectively 
short ranged, and the real part of $H$ resembles the ERE expansion. 
In the amplitude $T_{CS}$ the different strong and Coulomb coefficients 
proportional to a common power of $k$ can be grouped together, where 
one observes that they have comparable sizes. 
We therefore define 
\begin{equation}
\tilde{r}_0= r_0 -\frac{1}{3k_C}\, , 
\qquad \tilde{\cal P}_0={\cal P}_0\,+\,\frac{1}{15k_C^3}\,,
\label{eq:polepar2}
\end{equation}
and so on. Up to NLO we rewrite $T_{CS}$ as 
\begin{eqnarray}
T_{CS}&=&-\frac{2\pi}{\mu}\,\frac{C_{\eta}^2\,e^{2i\sigma_0}}
{-1/a_0+\tilde{r}_0k^2/2-\tilde{\cal P}_0 k^4/4-ikC_{\eta}^2}
\nonumber\\[1mm]&=&
-\frac{2\pi}{\mu}\,\frac{C_{\eta}^2\,e^{2i\sigma_0}}
{\tilde{r}_0(k^2-k_R^2)/2\!-\!\tilde{\cal P}_0(k^4\!-\!k_R^4)/4\!
-\!ikC_{\eta}^2}
\nonumber\\[1mm]&=&
-\frac{2\pi}{\mu}\,C_{\eta}^2\,e^{2i\sigma_0}\,\Bigg[
\underbrace{ 
\frac{1}{\tilde{r}_0(k^2-k_R^2)/2\!-\!ikC_{\eta}^2} }_{\rm LO\;\; term}
\nonumber\\&&
+\underbrace{ 
\frac{\tilde{\cal P}_0}{4}\,
\frac{(k^4\!-\!k_R^4)}
{(\tilde{r}_0(k^2-k_R^2)/2\!-\!ikC_{\eta}^2)^2}
}_{\rm NLO\;\; correction}+\ldots\,\Bigg],
\label{eq:poleTCS}
\end{eqnarray}
where
\begin{equation}
k_R^2 = \frac{2}{a_0 \tilde{r}_0}\,
\Bigg(\,\underbrace{1\raisebox{-11pt}{} }_{\rm LO\;\; term}
-\underbrace{\frac{\tilde{\cal P}_0}
{a_0 \tilde{r}_0^2}}_{\rm NLO\;\; correction}+\ldots\Bigg) \,.
\label{eq:polepos}
\end{equation}
From this expression one sees directly that $k_R\sim M_{lo}$, 
with corrections of ${\cal O}(M_{lo}^2/M_{hi}^2)$. 
Note that 
we keep the exact form of the imaginary term in Eq.~(\ref{eq:poleTCS}): 
even though it is negligible at $k\sim k_R$, 
it has an 
important exponential dependence on the energy 
responsible for keeping the phase shifts real in the elastic regime. 

When $a_0<0$ and $r_0< 1/3k_C$, we have $k_R^2>0$ and the two 
poles of Eq. (\ref{eq:poleTCS}) are located 
in the lower half of the complex-momentum plane very near the
real axis, as expected for a narrow resonance. 
The amplitude $T_{CS}$ can be written in terms of the resonance energy 
$E_R=k_R^2/2\mu$ and the resonance width $\Gamma(E)$ as\footnote{See 
also Ref.\cite{gelman}.}
\begin{equation}
T_{CS}=
\frac{2\pi e^{2i\sigma_0}}{\mu \sqrt{2\mu E}}\,\frac{\Gamma(E)/2}
{E-E_R+i \Gamma(E)/2}\,.
\label{eq:poleTCSwidth}
\end{equation}
One finds that
\begin{eqnarray}
&&\Gamma(E)=\Gamma(E_R) \frac{e^{2\pi k_C/k_R}-1}{e^{2\pi k_C/k}-1}
\Bigg[\,\underbrace{1\raisebox{-11pt}{} }_{\rm LO\;\; term}
\nonumber\\&&
-\underbrace{\frac{\mu^2 \tilde{\cal P}_0}{2\pi k_C} 
\left(e^{2\pi k_C/k_R}-1\right)
\frac{\Gamma(E_R)}{2} (E-E_R)\raisebox{-11pt}{} }_{\rm NLO\;\; correction}
+\ldots\Bigg]
\,,
\label{eq:widthstrongE}
\end{eqnarray}
where
\begin{equation}
\Gamma(E_R)
=-\frac{4\pi k_C}{\mu \tilde{r}_0} \frac{1}{e^{2\pi k_C/k_R}-1}
\Bigg(\,\underbrace{1\raisebox{-11pt}{} }_{\rm LO\;\; term}
+\underbrace{\frac{\tilde{\cal P}_0k_R^2}{\tilde{r}_0}}_{\rm NLO\;\; correction}
+\ldots\Bigg)
\,.
\label{eq:widthstrong}
\end{equation}
The width is very small because of the large value of $2\pi k_C/k_R$
in the exponential.

In the form of
Eqs. (\ref{eq:poleTCSwidth}) and (\ref{eq:widthstrongE})
we can keep
$E_R$
and $\Gamma(E_R)$ fixed at each order in the expansion.
Note that these equations
do not change to this order if one makes a different choice
---{\it e.g.}, $(k^2-k_R^2)^2$ instead of $k^4-k_R^4$---
for the form of the $\tilde{\cal P}_0$ term in Eq. (\ref{eq:poleTCS}).
The behavior of the phase shift around $E_R$ is guaranteed to be of a 
resonant type, as it automatically satisfies the constraints 
\begin{equation}
\delta_0^c(E_R)=\frac{\pi}{2}\quad\mbox{(resonance energy)}\,,
\label{eq:constr-en}
\end{equation}
and 
\begin{equation}
\left(\frac{d\delta_0^c(E)}{dE}\right)_{E_R}=\frac{2}{\Gamma(E_R)}
\quad\mbox{(resonance width)}\,.
\label{eq:constr-wd}
\end{equation}

\subsection{confronting the data}

Unfortunately $\alpha\alpha$ scattering data at low energies are not 
abundant. Nevertheless, all the existing measurements at 
$E_{LAB}$ up to $5$~MeV show that it is dominated by the $S$ wave, 
thanks to the presence of the ($J^\pi$, $I$) = ($0^+$, 0) resonance 
immediately above threshold. 
Early determinations of the $0^+$ resonance energy were performed in 
reactions like ${}^{11}{\rm B}+p\to 2\alpha+\alpha$ (see 
\cite{aaa69} and references therein). 
Later measurements of the scattering of $^4$He atoms on $^4$He$^+$ ions, 
especially projected to scan the resonance energy region and supplemented 
with a detailed analysis \cite{Ben68,Wue92} improved the determination 
of the resonance energy and width to their currently accepted 
values, $E_{LAB}^R=184.15\pm0.07$~keV and 
$\Gamma_{LAB}^R=11.14\pm0.50$~eV \cite{Wue92}. 
The resonance CM momentum is thus 
$k_R=\sqrt{\mu E_{LAB}^R}\approx 18.5$~MeV. 

The obvious way to extract the ERE parameters from the $S$-wave 
phase shift is to fit (the cotangent of) the latter to the ERE formula. 
However, fits to $\alpha\alpha$ scattering data alone are not able to 
constrain well those parameters and, apart from the large uncertainties, 
don't seem to predict the existence of the $0^+$ resonance. Imposing that 
the phase shift crosses the value $\pi/2$ at the resonance energy 
improves the situation, though not dramatically. 
Following a previous suggestion \cite{T66}, Ref. \cite{R67} used not 
only the available resonance energy, but also the width \cite{Ben68} 
to constrain these parameters. The obtained values for 
$a_0=-1.76\times~10^{3}$~fm, $r_0=1.096$~fm, and 
${\cal P}_0=-1.654$~fm${}^{3}$ made use of the resonance energy and 
width from Ref. \cite{Ben68}. Later we compare these numbers with 
the ones from our EFT fits. 

At energies below $E_{LAB}=3$~MeV data were obtained, and a phase-shift 
analysis performed, by Ref. \cite{HT56}. Values of the latter can be 
found in Table II of the review~\cite{aaa69}. 
Since it is well-determined experimentally, and due to its relevance to 
the triple-alpha process, we use the $0^+$ resonance parameters from 
Ref.~\cite{Wue92} as important constraints. 
This is in line with the EFT approach, where lower-energy observables 
have preference over higher-energy ones. 
The relationship among the EFT parameters and the resonance energy and 
width allows one to reduce the number of variables to be 
adjusted at each order in the power counting. 
Below, we also show the ERE from Ref. \cite{R67} for orientation, 
and comment on the extremely large 
value of the scattering length $a_0$, which suggests a 
large amount of 
fine-tuning in the parameters of the underlying theory away from the 
naturalness assumption. 

In the power counting that we discussed for the $\alpha\alpha$ 
system, the amplitude $T_{CS}$ for generic momenta is given 
up to NLO by Eq. (\ref{eq:poleTCS}). As demonstrated in the previous 
subsection, this expression combines the deep-non-perturbative 
Coulomb approximation (\ref{eq:Happrox}) for the function $H$ with 
the expansion around the resonance pole, thus 
preventing the need for multiple kinematical fine-tunings. 
In LO, the two parameters $a_0$ and $\tilde{r}_0$ can be obtained 
from the constraints (\ref{eq:polepos}) and (\ref{eq:widthstrong}). 
At NLO, a fit to scattering data is needed to determine $\tilde{\cal P}_0$. 

\begin{figure}[htb]
\begin{tabular}{c}
\includegraphics[width=3.3in]{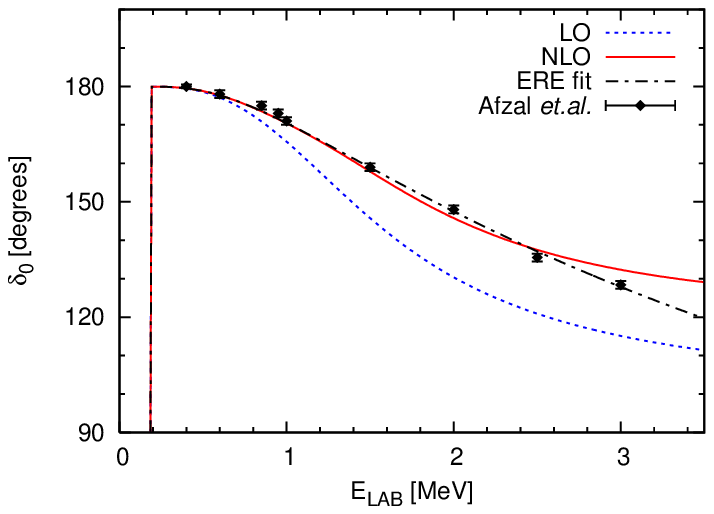}\\
\includegraphics[width=3.3in]{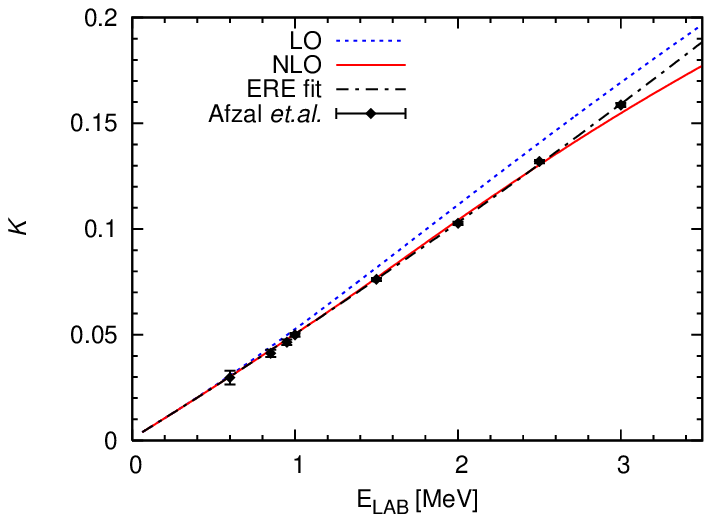}
\end{tabular}
\vspace*{8pt}
\caption{EFT results for $\alpha\alpha$ scattering at LO (dotted) and NLO 
(solid), compared against the data. Top panel: phase shift $\delta_0^c$. 
Bottom panel: $K(\eta)\equiv C_{\eta}^2(\cot\delta_0^c-i)/2\eta+H(\eta)$.
\protect\label{fig2}}
\end{figure}

Figure \ref{fig2} shows our results with the resonance position and 
width constraints, compared to the available $S$-wave phase shifts 
below $E_{LAB}=3$~MeV. 
In the region above the resonance, where scattering data are shown, 
the LO curve is a prediction, which is consistent with 
the first few points but then moves away from the data. 
The NLO curve has $\tilde{\cal P}_0$ as an extra parameter, which is 
determined from a global $\chi^2$-fit to scattering data shown. 
As expected from a convergent expansion, 
the description of the 
low-energy data improves with increasing order. 
At about 3 MeV and above, higher-order 
contributions are expected to be important, as suggested by 
the discussion on the relevant scales and manifest in 
the growing difference between the NLO curve and both LO curve
and data points.
Also shown are results from a fit using the conventional ERE formula,
Eq.~(\ref{eq:ere}), in order to stress the differences between this
and our EFT approach.

Table~\ref{tab:erepar} shows the values of ERE parameters used to 
produce the curves of Fig.~\ref{fig2}, and compare them 
with the values \cite{R67} obtained from effective range theory. 
At LO, our values for $a_0$ and $r_0$ are consistent with 
the ones from Ref. \cite{R67}, remarkably $r_0$. 
The NLO corrections, however, spoil this initial LO agreement. 
The reason for this deviation could be 
due to the way the width constraint was obtained in Ref.~\cite{R67}. 
Its Eq.~(4) reads 
\begin{equation}
\frac{dh}{dk^2}(\eta_R)-\frac{1}{\mu\Gamma(E_R)}\,
\frac{\pi}{e^{2\pi k_C/k_R}-1}
=\frac{1}{4k_C}\left(r_0-{\cal P}_0\,k_R^2\right),
\end{equation}
where $h(\eta)\equiv \mbox{Re}[H(\eta)]$.
Following 
this reference's procedure, we reproduced the quoted value of the 
width (6.4 eV) only when $dh(\eta_R)/dk^2$ 
was approximated by $1/12k_C^2$ [see Eq.~(\ref{eq:Happrox})]. 
However, that is equivalent to neglect the electromagnetic term of 
$\tilde{\cal P}_0$ in Eq.~(\ref{eq:widthstrong}), 
which is inconsistent since the strong piece contributes at the same order. 
Neglecting this term explains the agreement at LO and disagreement at 
NLO between our and Ref. \cite{R67} numbers for $r_0$. 
With a smaller $r_0$, and therefore larger (negative) $\tilde r_0$, one 
can also understand why Ref. \cite{R67} obtains a smaller $a_0$, as 
the product $\tilde r_0a_0$ is inversely proportional to the resonance 
momentum squared. By repeating the 
Ref. \cite{R67}'s procedure including the width constraint
consistently, we obtained essentially the same values as in our EFT
fits. This updated ERE fit is also shown in Table~\ref{tab:erepar}.

\begin{table}[htb]
\begin{center}
\begin{tabular} {|c||c|c|c|}
\hline
 & $a_0$ ($10^3$~fm) & $r_0$ (fm) & ${\cal P}_0$ (fm${}^{3}$)
\\ \hline\hline
LO & $-1.80$ & 1.083 & --- \\ \hline
NLO & $-1.92\pm 0.09$ & $1.098\pm 0.005$ & $-1.46\pm 0.08$ \\ \hline
ERE (our fit) & $-1.92\pm 0.09$ & $1.099\pm 0.005$ & $-1.62\pm 0.08$ \\ \hline
ERE~\cite{R67} & $-1.65\pm 0.17$ & $1.084\pm 0.011$ & $-1.76\pm 0.22$
\\ \hline
\end{tabular}
\end{center}
\caption{ERE parameters 
extracted from EFT fits 
in the first two orders, 
compared with values from two ERE fits, our own 
and Ref.~\cite{R67}'s. 
\label{tab:erepar}}
\end{table}

Our fits reveal both effective range 
$r_0\sim 1/(180\mbox{ MeV})$ and shape parameter 
${\cal P}_0\sim 1/(170\mbox{ MeV})^3$ scaling with powers of $M_{hi}$, 
in agreement with our {\it a priori} estimate. 
The relative errors in $a_0$ and $\tilde r_0$ at LO are estimated to 
be of the order of the EFT parameter expansion, 
$M_{lo}/M_{hi}\sim 1/7\approx 15$\%. 
At NLO, the main source of uncertainty comes from the precision of the 
most recent measurement of the resonance width \cite{Wue92}, lying 
between 4--5\%. 
The uncertainty in $\tilde{\cal P}_0$, given by the $\chi^2$-fit, 
is of the same order. 
Note that the small relative error in $r_0$ compared to the one 
in $\tilde r_0$ is due to the former being an order of magnitude larger 
than the latter, as we discuss in the next subsection. 
The NLO errors found here are a factor 
of two smaller than the ones obtained by Ref.~\cite{R67}. 

One should stress that the accurate value of the resonance 
width, $\Gamma_R=5.57\pm 0.25$~eV, imposes tight constraints on 
our fits, through $a_0$ and $r_0$. 
A significant improvement in our NLO fit and overall agreement with data 
is observed. But looking in detail, one sees that the theoretical 
curve is not able to cross the error bands of many scattering points 
below 3~MeV. This can be inferred from the $\chi^2$/datum $\simeq 4$. 
In principle, a better agreement should be achieved by an N${}^2$LO 
calculation. However, that would introduce an extra parameter that is mostly 
determined by the scattering data, and an expected agreement could mask any 
possible inconsistencies between the phase shifts and the 
resonance parameters. 
The high NLO $\chi^2$/datum suggests 
that they are not compatible with each other or, at least, 
one of them has overestimated precision. As a test we fitted both 
our NLO EFT and ERE expressions to the scattering data without the 
constraints from the resonance width. In these two cases, 
description of $S$-wave phase shifts is much better 
at the expense of an underpredicted resonance width, 
$\Gamma_R=4.9\pm 0.6$~eV with ERE and $\Gamma_R=2.87\pm 0.23$~eV with EFT. 
The ERE result is still consistent with the measured $\Gamma_R$ 
thanks to its large 
error bar. In EFT, where lower-energy data have higher priority, 
the discrepancy is amplified. 
The problem is even more pronounced if the fit is performed using data up to 
2.5~MeV instead of 3~MeV: the results 
$\Gamma_R=4.2\pm 0.6$~eV with ERE and $\Gamma_R=2.93\pm 0.34$~eV with EFT 
fall beyond the quoted experimental 
error bars. Oddly, this tendency continues as one lowers the upper limit 
in the fit. 
Reanalyses of the existing low-energy data or even new measurements 
seem necessary to resolve this discrepancy. 

\subsection{fine-tuning puzzle}

A surprising feature of the $\alpha\alpha$ system is the very large 
magnitude of $a_0$, even if compared 
to the large scattering length observed in the nucleon-nucleon system. 
The latter case is thought to be the outcome of a fine-tuning in the 
QCD parameters, giving rise to an anomalously low momentum scale. 
It is plausible to expect that this fine-tuning propagates to 
heavier systems. 
However, the enormous value of $a_0$ in the $\alpha\alpha$ 
is suggestive of a more delicate tuning, with electromagnetic 
interactions playing a crucial role. 

To better understand the puzzle it is worth looking in details at 
Eq.~(\ref{eq:aarenorm}). The Coulomb loop contributions, proportional to 
the curly brackets, scale as $M_{hi}^2/\mu$, given that 
$g^2/2\pi\sim M_{hi}/\mu^2$ and $2k_C\sim M_{hi}$. For the scaling 
$\Delta^{(R)}\sim M_{lo}^2/\mu$ to hold the scale-dependent parameter 
$\Delta(\kappa)$ must be of the same size of the Coulomb loops, and 
strongly cancel each other to produce a result $(M_{lo}/M_{hi})^2\sim 100$ 
times smaller. This amount of fine-tuning is necessary to obtain a 
resonance at the right position. Similar cancellation is observed for 
the $P$-wave resonance in $n\alpha$~\cite{BHvK2}. However, in the 
present case the fine-tuning is caused by a delicate balance between 
the strong and electromagnetic forces. 

This fine-tuning scenario becomes even more enigmatic by considering 
the width. The tiny value of 
$\Gamma_R\approx 6$ eV is intimately related to the resonance momentum, 
as can be immediately seen in Eq.~(\ref{eq:widthstrong}). That is simply 
an effect of the Coulomb repulsion---the resonance trapped inside the 
envelope formed by the strong plus Coulomb potential has to tunnel 
through a larger potential barrier as its energy gets smaller. The 
probability of tunneling is proportional to the resonance width, from 
where the smallness of $\Gamma_R$ follows. 
Despite the tiny value, the width has an associated scale of 
$4\pi k_C/\mu\tilde{r}_0$ which is quite large. A natural estimative 
for this quantity would be $\sim M_{hi}^2/\mu$, implying that 
$\tilde r_0\sim 1/M_{hi}$. However, from the width constraint one 
finds $\tilde r_0\sim M_{lo}/M_{hi}^2$, leading to an $r_0$ that 
roughly cancels against $1/3k_C$ with a remainder about 10\% smaller. 
From Eq.~(\ref{eq:polepos}) one sees that the extra fine-tuning in 
$\tilde r_0$ is responsible for an extra increase in the scattering 
length by a factor of $M_{hi}/M_{lo}$, becoming effectively 
$|a_0|\sim M_{hi}^2/M_{lo}^3$. 
It is remarkable, that if the 
strong forces generated an $r_0$ 11\% larger the ${}^{8}$Be ground 
state would be bound, with drastic consequences in the formation of 
elements in the universe (see also Ref.\cite{oberh}).

\section{$p\alpha$ scattering}\label{sec:pa}

The $p\alpha$ scattering at low energies is mostly dominated by the 
$P_{3/2}$ partial wave, due to the presence of a resonance at proton 
energies $E_p\approx 2$ MeV. $S$-wave also gives an important contribution 
for $E_p\leq 5$ MeV, which is the low-energy region we are interested in 
(see below). 
In the $P_{1/2}$ wave the phase shift varies smoothly 
up to $E_p=18$ MeV, suggesting the existence of a very broad resonance for 
proton energies between 8 and 15 MeV \cite{tunl}. 
However, to the order in the power counting and energies that we consider, 
it behaves as a typical higher-order effect. 
Higher wave phase shifts are less than a few fractions of degree below 
$E_p\approx 6$ MeV, where $D$-waves start being relevant. 
Therefore, only two partial waves contribute to the low-energy EFT for 
the $p\alpha$ system, up to and including NLO. 

The Lagrangian for nucleon and alpha particles interacting via $S_{1/2}$ 
and $P_{3/2}$ partial waves is written as \cite{RBvK,BHvK1,BHvK2} 
\begin{eqnarray}
&&{\cal L}_{N\alpha}=
\phi^{\dagger}\Bigg[i{D}_0+\frac{\vec{D}^2}{2m_{\alpha}}\Bigg]\phi
+N^{\dagger}\Bigg[i{D}_0+\frac{\vec{D}^2}{2m_N}\Bigg]N
\nonumber\\&&
+\varsigma_{0+}s^{\dagger}\Big[-\Delta_{0+}\Big]s
+\varsigma_{1+}t^{\dagger}\Bigg[i{D}_0
+\frac{\vec{D}^2}{2(m_{\alpha}+m_N)}-\Delta_{1+}\Bigg]t
\nonumber\\&&
-\frac{1}{4}\,F_{\mu\nu}F^{\mu\nu}
+\frac{g_{1+}}{2}\,\Big\{t^{\dagger}\vec S^{\dagger}\cdot
\Big[N\vec{D}\phi-(\vec{D} N)\phi\Big]+\mbox{H.c.}
\nonumber\\&&
-r\Big[t^{\dagger}\vec S^{\dagger}\cdot\vec{D}(N\phi)
+\mbox{H.c.}\Big]\Big\}
+g_{0+}\Big[s^{\dagger}N\phi+\phi^{\dagger}N^{\dagger}s\Big]
\nonumber\\[2mm]&&
+\varsigma_{0+}s^{\dagger}\left[i{D}_{0}
+\frac{\vec{D}^2}{2(m_{\alpha}+m_N)}\right]s
\nonumber\\&&
+g'_{1+}t^{\dagger}\Bigg[i{D}_0
+\frac{\vec{D}^2}{2(m_{\alpha}+m_N)}\Bigg]^2t\,,
\label{eq:pa-lag}
\end{eqnarray}
where $\phi$ and $N$ are the alpha and nucleon fields with masses 
$m_{\alpha}$ and $m_N$, respectively, and 
\begin{equation}
r=(m_{\alpha}-m_N)/(m_{\alpha}+m_N)\,.
\end{equation}
The auxiliary dimeron fields $t$ and $s$ couple to $N\alpha$ in $P_{3/2}$ 
and $S_{1/2}$ waves, with leading-order coupling constants $g_{1+}$ and 
$g_{0+}$, respectively. 
The $S_i$'s are $2\times 4$ spin-transition matrices between $J=1/2$ and 
$J=3/2$ total angular momentum states. 
The sign variables $\varsigma_{0+}$, $\varsigma_{1+}=\pm 1$ in front 
of the kinetic term of the dimeron fields are adjusted to 
reproduce the signs of the respective effective ranges. 
The gauge-covariant derivative is defined in the usual way 
in terms of the photon field $A_{\mu}$, 
\begin{equation}
D_{\mu}=\partial_{\mu}+ieZ\,\frac{1+\tau_3}{2}\,A_{\mu}\,,
\end{equation}
where $Z$ is the charge of the ``particle'' whose field the 
covariant derivative acts on, $\tau_3$ is the $z$-direction Pauli matrix 
acting in isospin space, and $A_{\mu}$ is the photon field. 

The power counting in essence consists of establishing how the size of 
the (renormalized) EFT coupling constants depend on the momentum 
scales present in the system one wishes to describe. The former have 
a direct relation to observables, namely, the ERE parameters. 
In the $p\alpha$ system one is interested in describing the scattering 
region around the $P_{3/2}$ resonance located at $k=k_p\sim M_{lo}$, 
where 
\begin{equation}
k_p=k_r-ik_i
\label{eq:pa-pole}
\end{equation}
is the (complex) resonance momentum in terms of 
real quantities $k_r$ and $k_i$. Table \ref{tab:pa-res} lists the 
resonance parameters extracted from the so-called extended $R$-matrix 
analysis used in Ref.~\cite{csoto97}, which indicates that 
$k_r\sim M_{lo}\sim 50$ MeV and $k_i\sim M_{lo}^2/M_{hi}\sim 10$ MeV. 
The suppression of $k_i$ relative to $k_r$ matches with narrow character 
of the $P_{3/2}$ resonance. 
As in the $\alpha\alpha$ case~\cite{HHvK}, $M_{hi}$ 
is set as the momentum required to excite the $\alpha$ core or the 
pion mass, both around 140 MeV. Our expansion parameter is of the order 
of 1/3, allowing us to expect convergent results for proton energies 
up to 5 MeV. 
\begin{table}
\caption{$P_{3/2}$ p$\alpha$ resonance parameters
from the extended $R$-matrix analysis of Ref. \cite{csoto97}.}
\label{tab:pa-res}
\begin{center}
\begin{tabular}{|c|c||c|c|}
\hline
$k_r$ (MeV) & $k_i$ (MeV) & $E_R$ (MeV) & $\Gamma_R/2$ (MeV) \\
\hline
\raisebox{0pt}[12pt][6pt]{}
$51.1$
& $9.0$
& $1.69$
& $0.61$
\\
\hline
\end{tabular}
\end{center}
\end{table}

\begin{table}
\caption{
$S_{1/2}$ and $P_{3/2}$ p$\alpha$ parameters extracted
from the ERE fit of  Ref. \cite{arndt73}.}
\label{tab:pa-ere}
\begin{center}
\begin{tabular}{|c|c|}
\hline
 $a_{0+}$ (fm) & $r_{0+}$ (fm) \\
\hline
\raisebox{0pt}[12pt][6pt]{}
$4.97\pm 0.12$
&$1.295\pm 0.082$\\
\hline
\end{tabular}
\begin{tabular}{|c|c|c|}
\hline
 $a_{1+}$ (fm${}^{3}$) & $r_{1+}$ (fm${}^{-1}$) & ${\cal P}_{1+}$ (fm) \\
\hline
\raisebox{0pt}[12pt][6pt]{}
$-44.83\pm 0.51$
&$-0.365\pm 0.013$
&$-2.39\pm 0.15$ \\
\hline
\end{tabular}
\end{center}
\end{table}

Table \ref{tab:pa-ere} shows the numerical values of the $S_{1/2}$ and 
$P_{3/2}$ ERE parameters taken from Ref.~\cite{arndt73}. As we can 
notice, while ${\cal P}_{1+}/4\sim r_{0+}/2\sim 1/M_{hi}$ are in 
agreement with the natural assumption, the quantities 
$a_{0+}\sim 1/M_{lo}$, $a_{1+}\sim 1/M_{lo}^3$, and $r_{1+}/2\sim M_{lo}$ 
are fine-tuned. To be consistent with these scalings the EFT couplings 
in Eq.~(\ref{eq:pa-lag}) must behave as 
\begin{equation}
\Delta_{1+}^{(R)}\sim \frac{M_{lo}^2}{2\mu}\,,\quad
\frac{g_{1+}^{(R)\,2}}{3\pi}\sim\frac{1}{\mu^2M_{lo}}\,,\quad
\mbox{and}\quad
\frac{g'_{1+}}{4}\sim\frac{\mu}{M_{lo}M_{hi}}
\label{eq:scales1}
\end{equation}
in the $P_{3/2}$ channel, and 
\begin{equation}
\Delta_{0+}^{(R)}\sim \frac{M_{hi}M_{lo}}{2\mu}\qquad
\mbox{and}\qquad
\frac{g_{0+}^2}{\pi}\sim\frac{M_{hi}}{\mu^2}
\label{eq:scales2}
\end{equation}
in $S_{1/2}$, where $\mu$ now stands for the $p\alpha$ reduced mass. 
In the latter case the situation is nearly identical to 
the proton-proton case~\cite{clbeft}: up to NLO the $S_{1/2}$ amplitude 
reads 
\begin{eqnarray}
T_{0+}&=&-\frac{2\pi}{\mu}\,\frac{C_{\eta}^{2}e^{2i\sigma_0}}
{-1/a_{0+}-2k_C H(\eta)}
\left[1 -\frac{r_{0+} k^2/ 2}{-1/a_{0+}-2k_C H(\eta)} \right]\,.
\nonumber\\
\label{eq:resum0+amp}
\end{eqnarray}

For $P$-wave interactions, Coulomb can be introduced in a 
straightforward way~\cite{RBvK}. The evaluation of Feynman diagrams and 
the necessary resummation results in 
\begin{eqnarray}
&&T_{1+}=-\frac{2\pi}{\mu}\,\frac{C_{\eta}^{(1)\,2}e^{2i\sigma_1}k^2
P_{1+}(\theta)}
{-1/a_{1+}+r_{1+}k^2/2-2k_C H^{(1)}(\eta)}
\nonumber\\[1mm]&&\times
\left[1
+\frac{{\cal P}_{1+} k^4/ 4}{-1/a_{1+}+r_{1+}k^2/2-2k_C H^{(1)}(\eta)}
\right]\,,
\label{eq:pa-1+amp}
\end{eqnarray}
where 
\begin{eqnarray}
C_{\eta}^{(1)\,2}&=&(1+\eta^2)C_{\eta}^{2}\,,
\\[1mm]
H^{(1)}(\eta)&=&k^2(1+\eta^2)H(\eta)\,,
\end{eqnarray}
and the variable 
$k_C=Z_{\alpha}Z_{p}\mu\alpha_{em}$ is adapted to the $p\alpha$ case. 
The $P_{3/2}$ projector is given by 
\begin{equation}
P_{1+}(\theta)=2\cos\theta+{\mbox{\boldmath $\sigma$}}
\cdot\mathbf{\hat n}\sin\theta\,,
\end{equation}
where $\theta$ is the angle between $\mathbf{k}$ and $\mathbf{k'}$ ---the 
initial and final momenta in the center-of-mass frame, respectively--- and 
$\mathbf{\hat n}=\mathbf{k}\times\mathbf{k'}/|\mathbf{k}\times\mathbf{k'}|$ 
is the unit vector normal to the scattering plane. 
In this form, the $T_{1+}$ amplitude requires the same kinematical 
fine-tuning discussed in Sect.~\ref{sec:aa-pole}. For pragmatical 
reasons, we therefore adopt the expansion around the resonance pole 
in this channel. 

\subsection{$P_{3/2}$ resonance pole expansion}

For practical applications, the more efficient way of expressing the 
EFT amplitude in the presence of low-energy resonances is to perform 
the resonance pole expansion. One nice feature is that, at every order 
in the power-counting, the amplitude maintains its resonant behavior 
at the expected position with the correct magnitude. 
It is straightforward to extend the ideas outlined in Sect.~\ref{sec:aa-pole} 
to the present case, with a resonance pole in the fourth quadrant of 
the complex momentum plane~\cite{RBvK}. 
With the pole given by Eq.~(\ref{eq:pa-pole}) one rewrites 
Eq.~(\ref{eq:pa-1+amp}) as 
\begin{eqnarray}
T_{1+}&=&-\frac{2\pi}{\mu}\,\frac{C_{\eta}^{(1)\,2}e^{2i\sigma_1}k^2
P_{1+}(\theta)}
{\frac{\tilde{r}_{1+}}{2}(k^2-k_p^2)-2k_C\left[
H^{(1)}\left(\frac{k_C}{k}\right)-H^{(1)}\left(\frac{k_C}{k_p}\right)
\right]}
\nonumber\\[1mm]&&\times
\left\{1 + \frac{{\cal P}_{1+} (k^2-k_p^2)^2/4}
{\frac{\tilde{r}_{1+}}{2}(k^2-k_p^2)-2k_C\left[
H^{(1)}\left(\frac{k_C}{k}\right)-H^{(1)}\left(\frac{k_C}{k_p}\right)
\right]}
\right\}
\,,
\nonumber\\
\label{eq:1+amppoleprime}
\end{eqnarray}
where
\begin{equation}
\tilde{r}_{1+}=-k_r\Bigg\{
\underbrace{2L_i}_{\rm LO}
-
\underbrace{2ik_i\,{\cal P}_{1+}}_{\rm NLO}
\Bigg\}
\end{equation}
and $L_i$, $L_r$ are defined at the pole position from 
\begin{equation}
2k_CH^{(1)}(k_C/k_p)=k_r^3\,L_r+2ik_r^2k_i\,L_i\,.
\end{equation}
The factors in front of $L_r,L_i\sim O(1)$ reflect the behavior of the 
Coulomb function $H^{(1)}(\eta)$ around the resonance. Notice that, 
as in the $\alpha\alpha$ amplitude, $T_{1+}$ is parametrized only in 
terms of $k_r$ and $k_i$ at LO, with the extra parameter ${\cal P}_{1+}$ 
appearing at NLO. 

The $P_{3/2}$ scattering length and effective range are given by 
\begin{eqnarray}
\frac{r_{1+}}{2}&=&
-k_r\Bigg[
L_i-\frac{k_r{\cal P}_{1+}}{2}
\left(1-\frac{k_i^2}{k_r^2}\right)\Bigg]
\nonumber\\&\simeq&
-k_r\Bigg[
\underbrace{L_i}_{\rm LO}
-
\underbrace{\frac{k_r{\cal P}_{1+}}{2}}_{\rm NLO}
\Bigg]\,,
\label{eq:pa-r1plus}
\\[1mm]
\frac{1}{a_{1+}}&=&
-k_r^3\Bigg[L_r+L_i\left(1-\frac{k_i^2}{k_r^2}\right)
-\frac{k_r{\cal P}_{1+}}{4}
\left(1+\frac{k_i^2}{k_r^2}\right)^2\Bigg]
\nonumber\\&\simeq&
-k_r^3\Bigg[
\underbrace{L_r+L_i}_{\rm LO}
-
\underbrace{\frac{k_r{\cal P}_{1+}}{4}}_{\rm NLO}
\Bigg]\,,
\label{eq:pa-a1plus}
\end{eqnarray}
and nicely show the scalings $1/a_{1+}\sim M_{lo}^3$ and 
$r_{1+}/2\sim M_{lo}$ that were expected from the numbers 
of Table~\ref{tab:pa-ere}. 

\subsection{results}

In order to test our power-counting assumptions, we fit the expressions 
for the $S_{1/2}$ and $P_{3/2}$ amplitudes to the corresponding phase 
shifts and compare the obtained ERE parameters with the ones in the 
literature. We used the numbers from Table 5 of Ref.~\cite{arndt73} in 
our fit. The results are shown in Table~\ref{tab:pa-fit1}. 
In our fits we have assumed that all the points have the same
statistical weight, since error bars were not provided.
However, we incorporated the EFT systematic error estimates by imposing
that the amplitude cannot be reproduced to a precision better than
$(k/2m_{\pi})^{n+1}$, where $k$ is the
CM momentum of the
${\rm p}\alpha$ system, $m_{\pi}$ our high-energy scale (the pion mass),
and $n$ the order in the EFT expansion.
The maximum and minimum values of each ERE parameter were determined by 
multiplying the amplitude by a factor of $1+x(k/2m_{\pi})^{n+1}$, where 
$x$ is a constant that varies between $-1$ and $1$. For the central value,
such constant is zero.
As one can see, our results are consistent with the ones shown in 
Table~\ref{tab:pa-ere} at NLO. At LO the central values are slightly 
off, nevertheless consistent with an expansion parameter of the order 
of 1/3. 

\begin{table}
\caption{
$S_{1/2}$ and $P_{3/2}$ p$\alpha$ ERE parameters extracted in our
fits at LO and NLO.}
\label{tab:pa-fit1}
\begin{center}
\begin{tabular}{|c||c|c|}
\hline
order & $a_{0+}$ (fm) & $r_{0+}$ (fm) \\
\hline
\raisebox{0pt}[12pt][6pt]{ LO } &
$7.4^{+8.0}_{-2.2}$
& --- \\
\hline
\raisebox{0pt}[12pt][6pt]{ NLO } &
$4.81^{+0.05}_{-0.21}$
& $1.7^{+1.3}_{-0.8}$ \\
\hline
\end{tabular}
\end{center}

\begin{center}
\begin{tabular}{|c||c|c|c|}
\hline
order & $a_{1+}$ (fm${}^{3}$)
& $r_{1+}$ (fm${}^{-1}$) & ${\cal P}_{1+}$ (fm) \\
\hline
\raisebox{0pt}[12pt][6pt]{ LO } &
$-58.0^{+11.0}_{-29.0}$ 
&
$-0.15^{+0.14}_{-0.09}$ 
& --- \\
\hline
\raisebox{0pt}[12pt][6pt]{ NLO } &
$-44.5^{+1.6}_{-0.1}$ 
&
$-0.40^{+0.04}_{-0.10}$ 
& $-2.8^{+1.0}_{-1.8}$
\\
\hline
\end{tabular}
\end{center}
\end{table}

In Table~\ref{tab:res2} are shown our fit results for the $P_{3/2}$ 
resonance parameters. One should stress that $k_r$ and $k_i$ are the 
variables determined from the fit, from which $a_{1+}$ and $r_{1+}$ 
can be extracted via Eqs.~(\ref{eq:pa-a1plus}) and (\ref{eq:pa-r1plus}). 
The results are consistent with the values in Table~\ref{tab:pa-res}, 
except for $k_i$ which is slightly off. That is probably due to the 
fact that it is a subleading effect compared to $k_r$, 
$k_i\sim M_{lo}^2/M_{hi}$, and to improve the agreement one needs to 
go one order higher. Nevertheless, the resonance energy and width are 
consistent with Ref.~\cite{csoto97}, which used the extended 
$R$-matrix analysis. The latter is based on the same concept that the 
resonance is a manifestation of a pole of the amplitude in the lower 
half of the complex energy plane. Both our and 
Ref.~\cite{csoto97} results disagree with the traditional 
$R$-matrix analysis, which are nowadays considered less reliable 
in extracting resonance properties~\cite{tunl}. 

\begin{table}
\caption{$P_{3/2}$  $p\alpha$ resonance parameters
extracted from our fits at LO and NLO.}
\label{tab:res2}
\begin{center}
\begin{tabular}{|c||c|c||c|c|}
\hline
$P_{3/2}$ & $k_r$ (MeV) & $k_i$ (MeV) & $E_R$ (MeV) & $\Gamma_R/2$ (MeV) \\
\hline
\raisebox{0pt}[12pt][6pt]{ LO } &
$50.6^{+1.2}_{-2.5}$
& $10.3_{-0.8}^{+1.4}$
& $1.64^{+0.09}_{-0.18}$
& $0.70^{+0.12}_{-0.09}$
\\
\hline
\raisebox{0pt}[12pt][6pt]{ NLO } &
$50.7^{+0.5}_{-0.6}$
& $9.40_{-0.10}^{+0.01}$
& $1.66^{+0.04}_{-0.04}$
& $0.63^{+0.01}_{-0.02}$
\\
\hline
\end{tabular}
\end{center}
\end{table}

\begin{figure}[htb]
\centerline{
\includegraphics[width=3.3in]{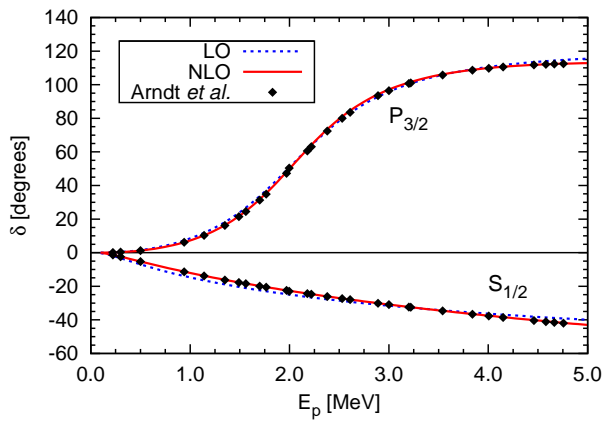}
}
\caption{EFT results for $S_{1/2}$ and $P_{3/2}$ scattering phase shifts 
at LO (dotted) and NLO (solid), compared against the partial wave analysis 
results from Arndt {\em et al.} 
(diamonds).}
\label{fig:pa-pshft}
\end{figure}

In Fig.~\ref{fig:pa-pshft} we show the EFT results for 
$S_{1/2}$ and $P_{3/2}$ phase shifts at LO and NLO. A very good convergence 
is observed, especially for the $P_{3/2}$ channel. The curves using 
the ERE expressions are omitted, since they essentially fall on top of 
the NLO curve and cannot be distinguished in a plot. 

\begin{figure}[htb]
\centerline{
\includegraphics[width=3.3in]{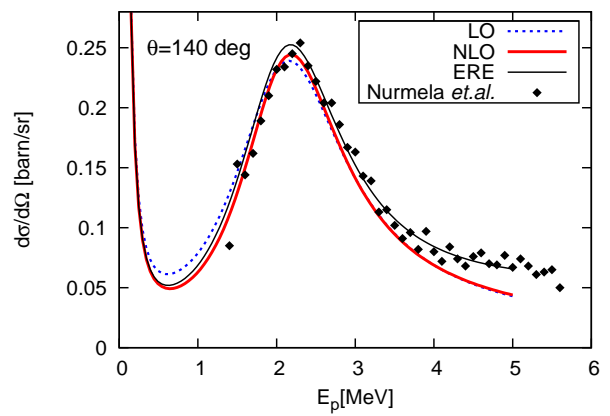}
}
\caption{EFT results at LO (dotted) and NLO (thick solid) for $p\alpha$ 
elastic cross-section at $\theta=140^{\circ}$ laboratory scattering angle, 
compared against the partial wave analysis results from Arndt {\em et al.} 
(thin solid) and measured data point from Ref.~\cite{nurmela} (diamonds).}
\label{fig:pa-xsec}
\end{figure}

Since we obtained a very good fit of the $S_{1/2}$ and $P_{3/2}$ EFT 
amplitudes to the corresponding phase shifts, it is useful to compare 
directly to observables in order to test our assumptions about the neglected 
higher-order terms. This is shown in Fig.~\ref{fig:pa-xsec} for the 
elastic differential cross-section measured at the laboratory scattering 
angle $\theta=140^{\circ}$. The sharp enhancement at low energies 
reflects the characteristic dominance of the Mott cross-section. 
One clearly sees that already at LO (dotted line) one has a good description 
of data, especially around the resonance peak. There is a slight improvement 
at NLO (thick solid line). The curves also agree well with the ERE curve 
(thin solid line), which uses the ERE formula for the $S_{1/2}$, $P_{3/2}$, 
and $P_{1/2}$ partial waves and the numerical parameters from 
Ref.~\cite{arndt73}. However, small deviations start to show up at 
$E_p\approx 3.5$ MeV, indicating an increasing importance of higher order 
terms as one goes higher in energy. The deviations, which seem to 
come mainly from the $P_{1/2}$ channel, can be eliminated by going 
to higher orders in the power counting. 

\section{summary}\label{sec:end}

I reviewed the programme of halo/cluster EFT suitable to study several 
interesting nuclear processes that play a role in nuclear astrophysics. 
I concentrated on two important aspects commonly present in those systems, 
namely, narrow resonances and Coulomb interactions. 
The $\alpha\alpha$ and $p\alpha$ interactions were given as examples 
of how setting up the formalism, then used as practical applications. 

The study of the $\alpha\alpha$ interaction revealed a very interesting 
picture, where the scales of the strong interaction conspire to produce 
a system with an almost non-relativistic conformal symmetry. One also 
observed extra amount of fine-tuning once electromagnetic interactions 
between the particles are turned on, up to the point of generating an 
$S$-wave scattering length that is three orders of magnitude larger 
than the typical range of the interaction. 
Despite the struggle in understanding the subtle cancellations between 
the strong and electromagnetic forces, the formalism is quite successful 
phenomenologically. In fact, thanks to those cancellations we were able 
to pin down 
the $S$-wave effective range parameters with an improved accuracy relative 
to previous studies. 

We extended the formalism to include $P$-waves with resonance and 
Coulomb interactions when dealing with the $p\alpha$ system. 
We performed the expansion of the $P_{3/2}$ amplitude around the 
resonance pole, which allowed us to extract the resonance properties 
directly from a fit to the phase shift. We derived expressions for 
the scattering length and effective range in terms of the resonance 
parameters, that explains the scalings of the former with the low and 
high momentum scales $M_{lo}$ and $M_{hi}$ of the theory. 
Our results at LO and NLO exhibit good convergence at a ratio of about 
1/3, and the resonance energy and width are consistent with the ones 
using the extended $R$-matrix analysis. Comparison with the differential 
cross-section at $140^{\circ}$ laboratory angle reassures the 
consistency of the power counting, with $P_{1/2}$ contribution starting 
to show up only for proton energies beyond 3.5 MeV. 

The $\alpha\alpha$ and nucleon-$\alpha$ interactions 
are the basic ones before considering more complicated clusters of $\alpha$ 
and nucleons. An interesting example is the Hoyle state in ${}^{12}$C, 
which plays a key role in the triple-alpha reaction responsible for the 
formation of heavy elements. 
A model-study with 
this state in mind was developed in Ref.\cite{limcyc}, where a perturbative 
treatment of the Coulomb interaction was proposed. This idea might be 
useful to handle the technical difficulties involving three charged 
particles. 

\section*{Acknowledgments}

I would like to thank Hans-Werner Hammer, Bira van Kolck, and Carlos
Bertulani for stimulating collaboration, and the organizers of the 
$19^{\rm th}$ International IUPAP Few-Body Conference 2009 for the 
opportunity to present this talk. 
This work was supported by the BMBF under contract number 06BN411.



\begin{thebibliography}{99}

\bibitem{bethe49} H. A. Bethe, Phys. Rev. {\bf 76}, 38 (1949).

\bibitem{biraere} U. van Kolck, Nucl. Phys. {\bf A645}, 273 (1999). 

\bibitem{CRS} 
J.-W. Chen, G. Rupak, and M.J. Savage, 
Nucl. Phys. {\bf A653}, 386 (1999); 
Phys. Lett. {\bf B464}, 1 (1999).

\bibitem{gautam} 
G. Rupak, Nucl. Phys. {\bf A678}, 405 (2000).

\bibitem{tho-efi} 
L.H.~Thomas,
Phys.\ Rev.\ {\bf 47}, 903 (1935); 
V. Efimov, Phys. Lett. {\bf 33B}, 563 (1970).

\bibitem{eftrev1} E.~Braaten and H.-W.~Hammer,
Phys.\ Rept.\  {\bf 428}, 259 (2006).

\bibitem{eftrev2} P.F.~Bedaque and U.~van Kolck,
Ann.\ Rev.\ Nucl.\ Part.\ Sci.\  {\bf 52}, 339 (2002).

\bibitem{BHvK00} P.F.~Bedaque, H.-W. Hammer, and U.~van Kolck,
Nucl. Phys. {\bf A676}, 357 (2000).

\bibitem{platter06} L. Platter, Phys. Rev. C {\bf 74}, 037001 (2006).

\bibitem{PHM05} L. Platter, H.-W. Hammer, and U.-G. Mei\ss ner, 
Phys. Lett. {\bf B607}, 254 (2005).

\bibitem{platterrev} L. Platter,
Few Body Syst. {\bf 46}, 139 (2009).

\bibitem{birse} M. C. Birse, J. A. McGovern, and K. G. Richardson, 
Phys. Lett. {\bf B464}, 169 (1999).

\bibitem{CH1} D. L. Canham and H.-W. Hammer, 
Eur. Phys. J. A {\bf 37}, 367 (2008). 

\bibitem{amorimetal} A.E.A. Amorim, T. Frederico, and L. Tomio,
Phys. Rev. C {\bf 56}, R2378 (1997);
L. Tomio, M.T. Yamashita, and T. Frederico,
Mod. Phys. Lett. A {\bf 24}, 998 (2009).

\bibitem{CH2} D. L. Canham and H.-W. Hammer, arXiv:0911.3238v1 [nucl-th].

\bibitem{BHvK1} C. A. Bertulani, H.-W. Hammer, and U. van Kolck,
Nucl. Phys. {\bf A712}, 37 (2002).

\bibitem{BHvK2} P. F. Bedaque, H.-W. Hammer, and U. van Kolck,
Phys. Lett. {\bf B569}, 159 (2003).

\bibitem{HHvK} R. Higa, H.-W. Hammer, and U. van Kolck, 
Nucl. Phys. {\bf A809}, 171 (2008).

\bibitem{clbeft} X.\ Kong and F.\ Ravndal,
Phys. Lett. {\bf B450} 320 (1999);
Nucl.\ Phys. {\bf A665}, 137 (2000).

\bibitem{aaa69} S.A. Afzal, A.A.Z. Ahmad, and S. Ali,
Rev. Mod. Phys. {\bf 41}, 247 (1969).

\bibitem{msw} T. Mehen, I.W. Stewart, and M.B. Wise, 
Phys. Lett. {\bf B474}, 145 (2000).

\bibitem{pole} D.R. Phillips, G. Rupak, and M.J. Savage,
Phys. Lett. {\bf B473}, 209 (2000).

\bibitem{gelman} B.A. Gelman, 
Phys. Rev. C {\bf 80}, 034005 (2009).

\bibitem{Ben68}
J. Benn, E.B. Dally, H.H. M\"uller, R.E. Pixley,
H.H. Staub, and H. Winkler,
Nucl. Phys. {\bf A106}, 296 (1967).

\bibitem{Wue92} 
S. W\"ustenbecker, H.W. Becker, H. Ebbing, W.H. Schulte,
M. Berheide, M. Buschmann, C. Rolfs, G.E. Mitchell, and J.S. Schweitzer,
Z. Phys. A {\bf 344}, 205 (1992).

\bibitem{T66} T.A. Tombrello, Phys. Lett. {\bf 23}, 106 (1966).

\bibitem{R67} G. Rasche, Nucl. Phys. {\bf A94}, 301 (1967).

\bibitem{HT56} N.P. Heydenburg and G.M. Temmer,
Phys. Rev. {\bf 104}, 123 (1956). 

\bibitem{oberh} H. Oberhummer, A. Cs\'ot\'o, and H. Schlattl,
Science {\bf 289}, 88 (2000).

\bibitem{tunl} D.R. Tilley, C.M. Cheves, J.L. Godwin, G.M. Hale, 
H.M. Hofmann, J.H. Kelley, C.G. Sheu, and H.R. Weller, 
Nucl. Phys. {\bf A708}, 3 (2002). 

\bibitem{RBvK} 
R. Higa, C.A. Bertulani, and U. van Kolck, in preparation. 

\bibitem{csoto97} A. Cs\'ot\'o and G.M.\ Hale,
Phys.\ Rev.\ C {\bf 55}, 536 (1997).

\bibitem{arndt73} R.A.\ Arndt, D.L.\ Long, and L.D.\ Roper,
Nucl.\ Phys.\ A {\bf 209}, 429 (1973).

\bibitem{nurmela} A. Nurmela, E. Rauhala, and J. R\"ais\"anen,
J. Appl. Phys. {\bf 82}, 1983 (1997). 

\bibitem{limcyc} H.-W. Hammer and R. Higa, 
Eur. Phys. J. A {\bf 37}, 193 (2008).

\end{thebibliography}
\end{document}